# Mode Properties of Flat-top Silver Nano-ridge Surface Plasmon Waveguides

Zeyu Pan,[1] Junpeng Guo,[1,*] Richard Soref,[2] Walter Buchwald,[3] and Greg Sun[2]

[1]Department of Electrical and Computer Engineering, University of Alabama in Huntsville, Huntsville, AL 35899

[2]Department of Physics, University of Massachusetts at Boston, Boston, MA 02125

[3] Solid State Scientific Corporation, Hollis, NH 03049

*Corresponding author: guoj@uah.edu

We investigate surface plasmon modes supported by flat-top silver nano-ridges. We calculate the mode electromagnetic field distribution, the dispersion curve, the travel range, and the figure-of-merit of the nano-ridge mode. We find that the nano-ridge surface plasmon modes are quasi-TEM modes with longitudinal field components three orders of magnitude smaller than the transverse field components. The quasi-TEM nature of mode profiles reveals that the propagation of free electron oscillations on the top of the nano-ridge contributes mainly to the tightly confined ridge mode. We also find that as the width of the nano-ridge decreases, the ridge mode becomes more tightly confined on the ridge top. As the width of the nano-ridge increases, the nano-ridge mode approaches two decoupled right-angle wedge plasmon modes.

OCIS codes: 240.6680, 230.7370.

# 1. Introduction

Surface plasmons are free electron density oscillations on the surface of metals in contact with dielectric materials [1-3]. Surface plasmons can propagate along the metal-dielectric boundaries, and form surface plasmon waves. Because surface plasmon waveguides can provide tightly confined sub-wavelength modes, surface plasmon modes in various waveguide structures, such as thin metal films [4-9], finite width thin film metal stripes and metal wires [10-19], trenches in metal surfaces [20-35], metal dielectric layer structures [36-42], dielectric-loaded metal films [43-54], and metal wedges [34, 35, 55-60] have been extensively investigated in the past. Recently, a round-top gold nano-ridge was fabricated using the focused ion beam milling [58] and that it was confirmed that metal nano-ridges can support surface plasmon modes. In this paper, we report our in-depth numerical investigations on the mode properties of flat-top silver nano-ridge waveguides. We calculate mode field distributions of the nano-ridge surface plasmon mode, the dispersion relation, propagation distance, mode size, and the figure-of-merit. We find that, compared with the metal wedge structures, flat-top nano-ridges have longer travel ranges and larger figure-of-merits. Our analysis in this paper provides a comprehensive view of the flat-top nano-ridge surface plasmon waveguides.

# 2. Surface plasmon mode of flat-top silver nano-ridges

The structure of the flat-top metal nano-ridge surface plasmon waveguide is shown in Fig. 1. The nano-scale metal ridge is extended in the $z$-direction. The surface plasmon wave propagates along the ridge top in the $z$-direction. The nano-ridge width ($w$) is in the $x$-direction. We assume the height of ridge is large that the substrate does not influence the ridge surface plasmon mode.

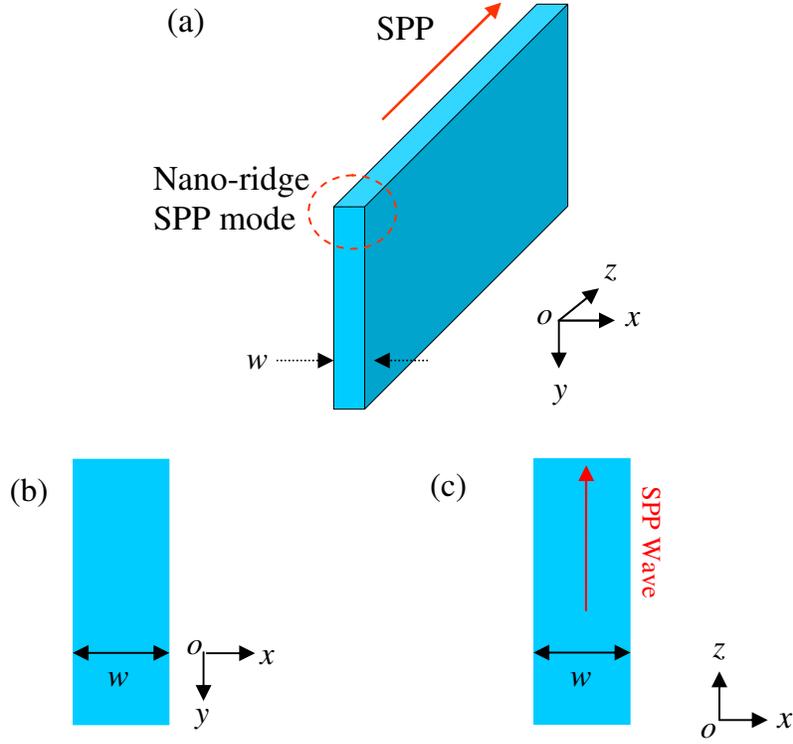

Fig. 1. (a) The 3D view of a metal nano-ridge plasmon waveguide. (b) The cross section of the flat-top nano-ridge plasmon waveguide. (c) The top view of the flat-top nano-ridge plasmon waveguide.

We calculate the guided surface plasmon-polariton (SPP) mode for a 120 nm wide silver nano-ridge at the telecommunication wavelength of 1.55 micron using a commercial mode solver (*Lumerical Solutions, Inc.*). The dielectric material around the silver ridge is air ($\varepsilon_d = 1.0$). The silver's electric permittivity at the 1.55 micron wavelength is $\varepsilon_m = -127.5 - 5.3j$ [61]. The calculated mode index is $n_{eff} = 1.018 - 0.00074j$. The attenuation coefficient is 261.87dB/cm. Fig. 2 (a)-(c) are the mode profiles of the three electric field components ($E_x$, $E_y$, $E_z$) of the 120 nm wide flat-top silver nano-ridge plasmon waveguide at the 1.55 micron wavelength. Fig. 2 (d)-(f) are the mode profiles of the three magnetic field components ($H_x$, $H_y$, $H_z$) of the flat-top nano-ridge waveguide at the same wavelength. The major components of the electric field are the transverse components, $E_x$ and $E_y$. The longitudinal component of the electric field $E_z$ is three orders of magnitude less than the transverse components. The major magnetic field components are also the transverse components, $H_x$ and $H_y$. The longitudinal component of the magnetic filed $H_z$ also is three orders of magnitude less than the transverse magnetic components. Therefore, nano-ridge surface plasmon modes can be considered as a quasi-transverse electromagnetic (TEM) mode. From Fig. 2, we can see that $E_y$ mode profile and $H_x$ mode profile are symmetrical with respect to the center of the metal ridge (i.e. $x = 0$ plane), while, the $E_x$ and $H_y$ profiles are anti-symmetrical with respect to the $x = 0$ plane. The symmetric properties of the mode field profiles are consistent with what we expect for the surface plasmon-polariton mode propagating along the top surface of the metal ridge. The nano-ridge plasmon mode can be considered as the two coupled 90° wedge plasmon modes. At the 1.55 micron wavelength, the mode index of a 90° silver wedge plasmon waveguide is 1.00675-0.000307*j*, the corresponding attenuation coefficient is 108.30 dB/cm. Compared with the 90° wedge mode, the 120 nm wide nano-ridge mode has a larger real part of the mode index. We will show later that this is true for all flat-top ridge waveguides of different widths.

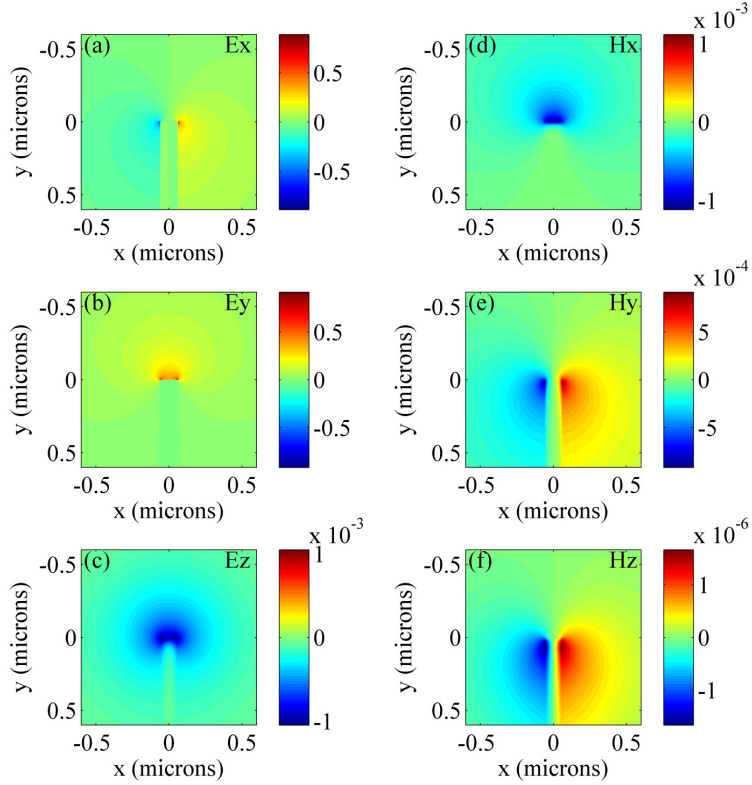

Fig. 2. (a)-(c) are the electric field ($E_x$, $E_y$, $E_z$) mode profiles of a 120 nm wide flat-top silver nano-ridge at 1.55 micron wavelength; (d)-(f) are the magnetic field ($H_x$, $H_y$, $H_z$) mode profiles of a 120 nm wide flat-top silver nano-ridge at 1.55 micron wavelength.

We calculated the mode dispersion curves of flat-top silver nano-ridge plasmon waveguides of different ridge widths. We also calculated the dispersion curves of the silver-air flat boundary and the right-angle (90°) silver wedge plasmon modes. The calculated dispersion curves are shown in the Fig. 3. The black solid line, black dashed line, red dashed line are the dispersion curves of the light line in the air, the flat silver-air interface surface plasmon mode, and the 90° wedge plasmon mode, respectively. As the frequency increases, the dispersion curve moves away from the light line, suggesting the reduction of the phase velocity and tighter mode confinement. When the flat-top nano-ridge width is increased, the dispersion curve of the nano-ridge mode moves toward the light line, approaching the dispersion curve of the 90° wedge plasmon mode. This can be explained that when the width of the metal ridge becomes wide, the two coupled wedge modes become decoupled.

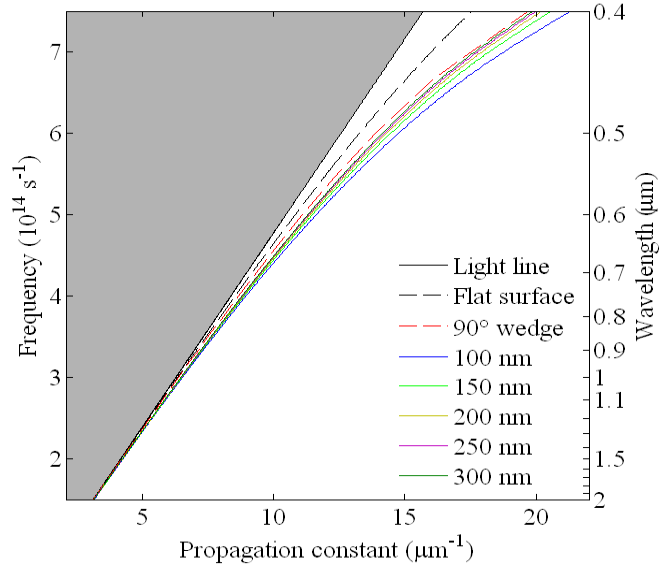

Fig. 3. The dispersion curves of silver flat-top nano-ridge plasmon waveguides of different ridge widths and the dispersion curves of the right-angle wedge plasmon mode and the flat surface plasmon mode.

Figure 4(a) shows the real part of the mode index versus the wavelength for several different ridge widths. The black dashed line in the Fig. 4(a) shows the real part of the mode index of the silver-air flat boundary surface plasmon mode. The red dashed line in Fig. 4(a) shows the real part of the mode index of the 90° wedge waveguide mode. It can be seen in Fig. 4(a) that when the ridge width increases, the real part of the mode index decreases, suggesting the decrease of the mode confinement. Fig. 4(b) shows the imaginary part of the mode index versus wavelength for different the ridge widths. The black dashed line in Fig. 4(b) shows the imaginary part of the mode index of the silver-air flat surface plasmon waveguide. The red dashed line represents the imaginary part of the mode index of the 90° wedge waveguide mode. It can be seen that when the width of the flat-top ridge increases, the imaginary part of the mode index decreases, indicating the reduction of the propagation loss.

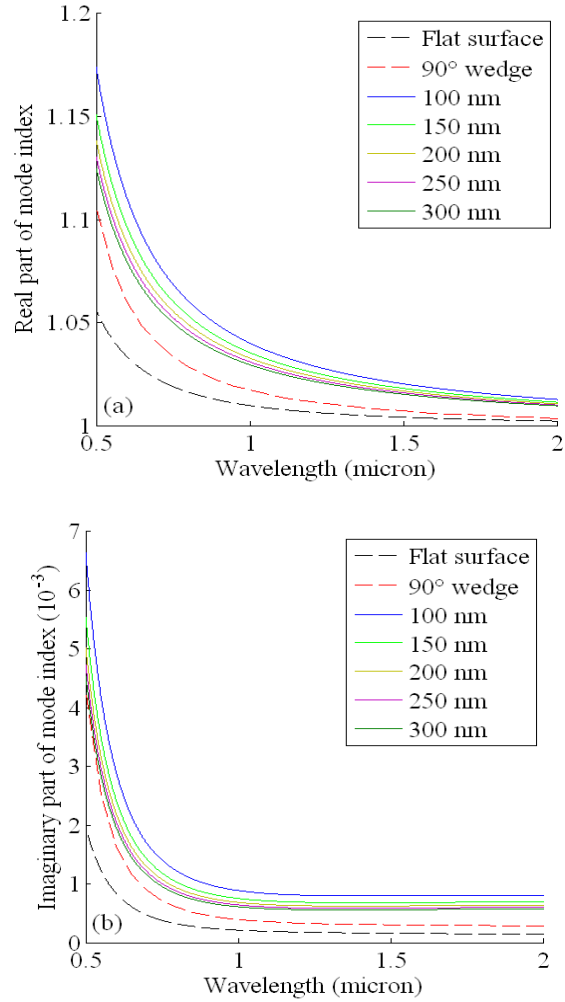

Fig. 4. (a) The real part of the mode index versus the wavelength for different ridge widths. (b) The imaginary part of the mode index versus the wavelength for different ridge widths.

We also calculated the real part and the imaginary part of the mode index versus ridge width at the wavelength of 1.55 micron. The results are shown in Fig. 5. As the ridge width increases, both the real part and the imaginary part are reduced. This indicates when the ridge becomes wider, the confinement becomes less and the attenuation becomes smaller. The real and imaginary parts of the mode index of a 90° silver wedge waveguide are also shown in the Fig. 5. The black and blue dashed lines are the real and imaginary part of the 90° wedge mode, respectively. We have discussed earlier in this paper that the nano-ridge plasmon mode can be considered as the hybrid mode of two coupled 90° wedge modes. When the width of the nano-ridge increases, the coupling between the two 90° wedge modes becomes weaker. Therefore, the ridge mode approaches the two separate 90° wedge modes.

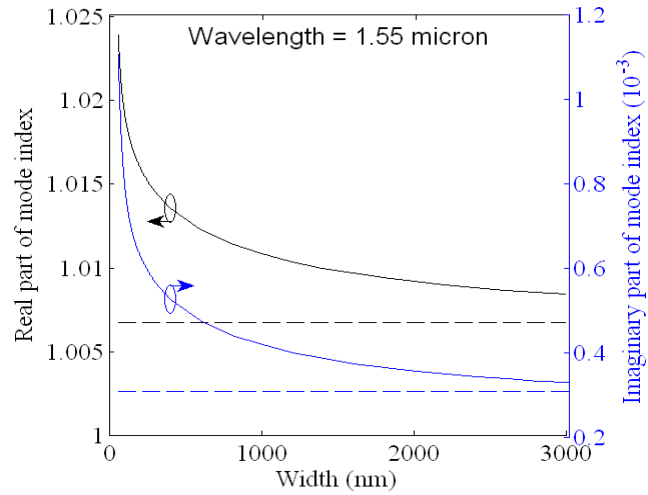

Fig. 5. The real and imaginary part of the mode index as the function of the ridge width at 1.55 micron wavelength.

The complex wave vector along the direction of the propagation is $\beta_z = n_{eff} k_0 = \beta - i\alpha$, where $\beta$ is the phase propagation constant of the mode, and $\alpha$ is the attenuation constant. The travel range is defined as the distance when the mode intensity attenuates to $1/e$ of its initial value, i.e. $L_p = 1/(2\alpha)$. Fig. 6 shows the travel range versus the free space wavelength and ridge width for the silver flat-top nano-ridge waveguide. Here it can be seen that as the ridge width increases, the travel range also increases.

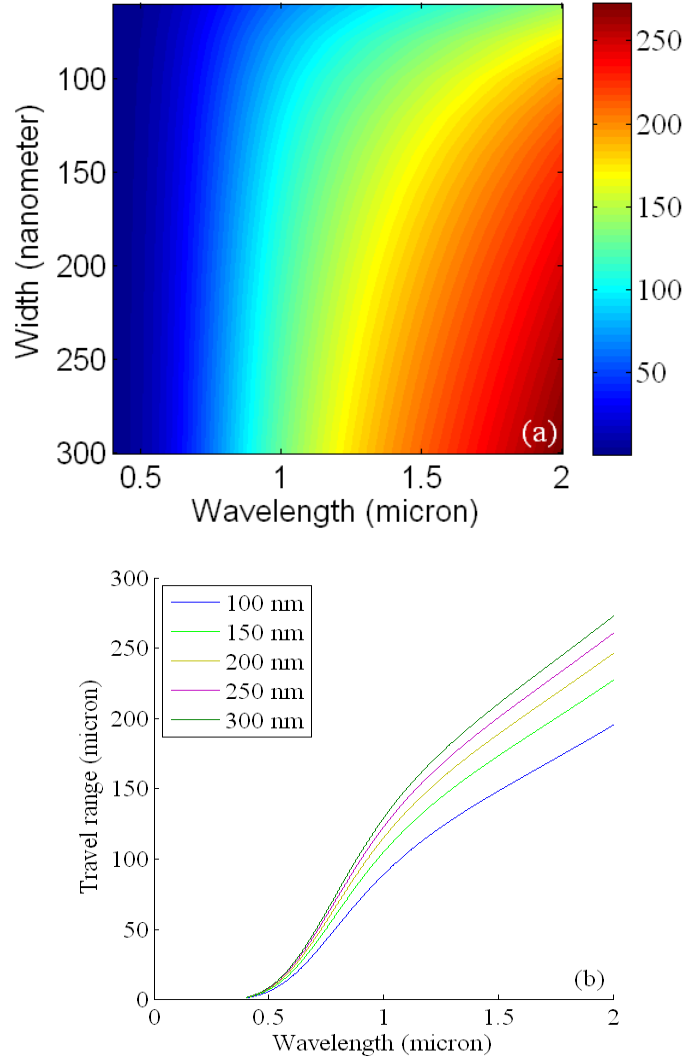

Fig. 6. (a) The travel range versus the wavelength and ridge width. (b) The line plot of the travel range of the flat-top nano-ridge plasmon mode as the function of free space wavelength for different ridge widths.

From the mode calculations, we can obtain the complex propagation constant of the ridge plasmon mode: $\beta_z = \beta - i\alpha$, where $\beta$ is the phase propagation constant, and $\alpha$ is the attenuation constant, both in the direction of propagation. Outside the metal in the surrounding dielectric medium, the transverse component of the complex wave vector is $\beta_\perp = \gamma - i\delta$ where $\gamma$ and $\delta$ describe the field oscillation and the decay in the transverse direction, respectively. From Maxwell equations, the complex propagation constants in the propagation direction and the transverse directions are related as

$$(\beta - j\alpha)^2 + (\gamma - j\delta)^2 = \varepsilon_d k_o^2 \tag{1}$$

where $k_0$ is the propagation constant in the free space, and $\varepsilon_d$ is the dielectric constant of the surrounding dielectric. Solving equation (1), we can obtain:

$$\gamma = -\left\{ \frac{\sqrt{[\beta^2 - \alpha^2 - \varepsilon_d k_o^2]^2 + 4(\alpha\beta)^2} - [\beta^2 - \alpha^2 - \varepsilon_d k_o^2]}{2} \right\}^{\frac{1}{2}} \tag{2}$$

$$\delta = \left\{ \frac{\sqrt{[\beta^2 - \alpha^2 - \varepsilon_d k_o^2]^2 + 4(\alpha\beta)^2} + [\beta^2 - \alpha^2 - \varepsilon_d k_o^2]}{2} \right\}^{\frac{1}{2}} \tag{3}$$

With the calculated $\delta$, we can find the mode size of the ridge mode. Here we define the mode size as $1/(2\delta) + w + 1/(2\delta) = w + 1/\delta$. We calculated the mode size of different width ridge plasmon waveguides at different free space wavelengths. The results are shown in the of Fig. 7. Fig. 7(a) shows the mode size versus the free space wavelength and the ridge width. We can see that as the wavelength or the ridge width increases, the mode size increases. Fig. 7(b) shows the mode size versus the free space wavelength for flat-top silver ridges of different ridge widths.

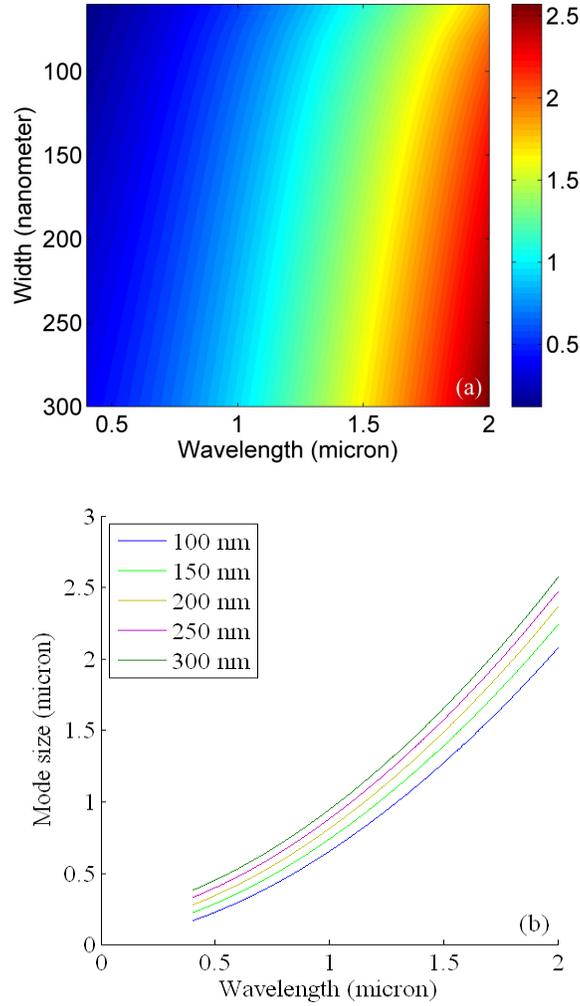

Fig. 7. (a) The mode size versus the wavelength and ridge width. (b) The mode size of the flat-top nano-ridge plasmon mode versus the free space wavelength for different flat-top ridges.

There is always a trade-off between the propagation attenuation and the mode confinement for surface plasmon waveguides [13, 62]. While the small mode confinement is the merit, the attenuation is the cost. The trade-off between the mode confinement and the propagation attenuation also holds for nano-ridge plasmon waveguides. Figure-of-merits of plasmon waveguides were proposed to characterize the trade-off between the attenuation and the confinement [63, 64]. Here, we define the figure-of-merit of the flat-top ridge waveguide as the ratio of the propagation distance over the mode size

$$FoM = (1/2\alpha)/(w+1/\delta). \tag{4}$$

We calculated the figure-of-merit for the nano-ridge waveguide versus the wavelength and the ridge width. The results are shown in the Fig. 8(a). Fig. 8(b) shows the line plots of the figure-of-merit versus the wavelength for several different ridges of different widths. It can be seen that the figure-of-merit reaches the maximum at 1.1 micron wavelength independent of ridge width. As the wavelength increases, the figure-of-merit first increases and then decreases.

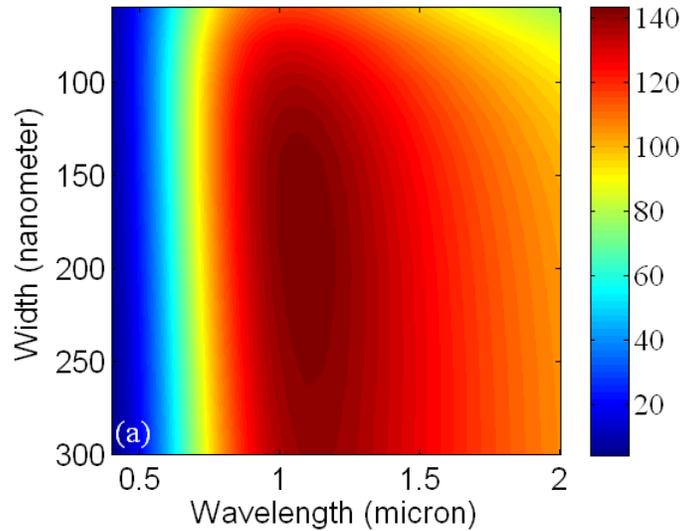

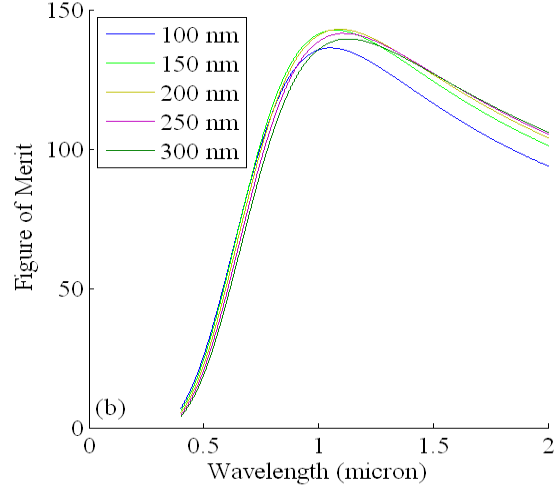

Fig. 8. (a) The figure-of-merit versus the wavelength and the ridge width. (b) The line plot of the figure-of-merit versus the wavelength for several ridge widths.

It might be due to the optical properties of the silver metal and the surrounding dielectric that causes the figure-of-merit peak around the 1.1 micron wavelength. This suggests 1.1 micron as being the optimal operational wavelength for silver nano-ridge waveguides embedded in the air. To compare the mode properties at the 1.1 micron wavelength and at the 1.55 micron wavelength, we calculated the travel range and the figure-of-merit of the silver nano-ridge waveguide versus the ridge width at these two wavelengths. The results are shown in Fig. 9(a) and (b) respectively. As the width of the nano-ridge increases, the travel range increases as expected. Further, the figure-of-merit is also seen to increase and peaks at a maximum value of roughly 143 for the 1.1 micron wavelength case, and reaches a plateau of about 124 for the 1.55 micron wavelength case. At both wavelengths, maximum values for the figure-of-merit occur in a ridge width range between 120 nm and 300 nm, suggesting an optimal range of the ridge width to address the trade-off between the propagation distance and the mode confinement.

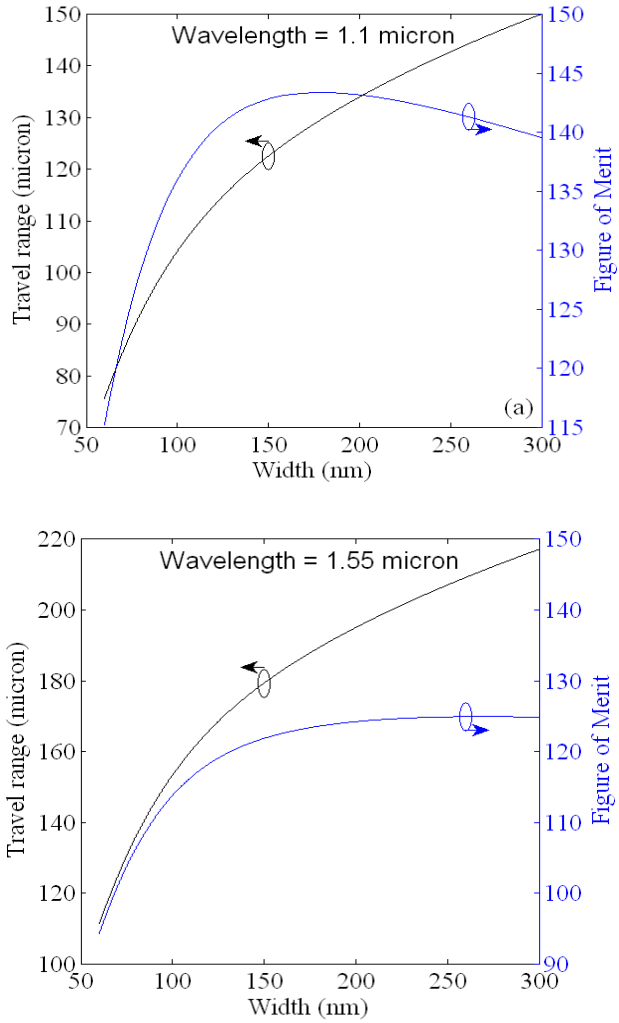

Fig. 9. The travel range and figure-of-merit of ridge mode versus the nano-ridge width at (a) 1.1 micron wavelength, (b) 1.55 micron wavelength.

## 3. Summary

We investigated the plasmon mode supported by flat-top silver nano-ridge waveguides. We have shown that the nano-ridge plasmon modes are quasi-TEM modes with very small longitudinal electromagnetic field components. The mode field profiles reveal that the propagation of the free electron density oscillations on the top of the nano-ridge contributes to the tightly confined ridge surface plasmon mode. We discussed that the nano-ridge mode can be considered as a hybrid mode of two 90° wedge plasmon modes.

When the ridge width increases and becomes wide, the nano-ridge mode approaches two decoupled 90° wedge plasmon modes. We calculated the dispersion, mode index, travel range, and the mode size of nano-ridge waveguides of different widths. We introduced the figure-of-merit for the nano-ridge plasmon waveguide to address the trade-off between the mode confinement and the travel range. We find that the figure-of-merit reaches a maximum at 1.1 micron wavelength, slightly depending on the width of the ridge waveguide. The 170 nm wide silver nano-ridge gives the highest figure-of-merit at 1.1 micron wavelength. At the 1.55 micron wavelength, the figure-of-merit reaches a plateau at the nano-ridge width of 180 nm, which gives a travel range 189.3 micron. For building plasmonic circuits, a ridge width between 120 nm and 300 nm represents a reasonable trade-off between confinement and travel range. Although we investigated the plasmon modes guided by silver metal nano-ridges, and the ridges can be other types of high electron density materials such as heavily doped semiconductors for infrared wavelength range [65, 66]. Flat-top nano-ridge surface plasmon waveguides are easy to fabricate. Therefore, they can be very useful as building blocks for realizing integrated plasmonic circuits at different wavelength regimes.

## Acknowledgment

Junpeng Guo acknowledges the support from the ASEE-Air Force Summer Faculty Research Fellowship Program. This work was also partially supported by the National Science Foundation (NSF) through the grant NSF-0814103 and the National Aeronautics and Space Administration (NASA) through the contract NNX07 AL52A.